\documentclass[preprint]{aastex631}
\usepackage{natbib}
\newcommand{\msunyr}{\ensuremath{{\rm M}_\odot ~ {\rm yr}^{-1}}}
\newcommand{\kms}{km\,s$^{-1}$}

\begin{document}

\title{Merger Interactions Enhance Star Formation Rates in Galaxy Group Housing QSO  PKS0405$-$123 and Gaseous Nebulae}
\shorttitle{Merger Interactions Enhance SFRs in Galaxies}
\shortauthors{Wolter et al.}


\author[0000-0003-4103-3186]{Ian E. Wolter}
\affiliation{Department of Astronomy, Smith College, Northampton, MA 01063, USA}
\affiliation{Department of Astronomy, The University of Texas at Austin, Austin, TX 78712, USA}

\author[0000-0002-8518-6638]{Michelle A. Berg}
\affiliation{Department of Astronomy, The University of Texas at Austin, Austin, TX 78712, USA}

\author[0000-0002-0302-2577]{John Chisholm}
\affiliation{Department of Astronomy, The University of Texas at Austin, Austin, TX 78712, USA}


\begin{abstract} 

The star formation rate (SFR) of galaxies can change due to interactions between galaxies, stellar feedback ejection of gas into the circumgalactic medium, and energy injection from accretion onto black holes. However, it is not clear which of these processes dominantly alters the formation of stars within galaxies. \citet{johnson2018} reported the discovery of large gaseous nebulae in the intragroup medium of a galaxy group housing QSO PKS0405$-$123 and hypothesized they were created by galaxy interactions. We identify a sample of 30 group member galaxies at z$\sim$0.57 from the VLT/MUSE observations of the field and calculate their [OII]$\lambda$$\lambda$3727,3729 SFRs in order to investigate whether the QSO and nebulae have affected the SFRs of the surrounding galaxies. We find that star formation is more prevalent in galaxies within the nebulae, signifying galaxy interactions are fueling higher SFRs.


\end{abstract}



\section{Introduction} 

The circumgalactic  medium (CGM) describes the gas surrounding a galaxy that is outside of the disk, but within the bounds of the virial radius \citep{Tumlinson_2017}. 
The relative structure of the CGM holds important clues for the integrated impact of galaxies and AGN. A feasible driver of QSO fueling and star formation is galaxy interactions, which exchange gas and dust in the CGM.
However, these processes can also slow down the conversion of gas into stars through heating up the gas and/or removing cool fuel from the CGM.

\citet{johnson2018} presents the discovery of several large ionized gaseous nebulae in the intragroup medium of galaxies surrounding QSO PKS0405$-$123 (their Figure 2). The three largest nebulae sit to the east, south, and east-by-south east of the QSO. In this research note, we characterize the SFRs for a sample of 30 galaxies within the intragroup medium in order to assess the effects the presence of these large sources of potential fuel and a QSO have on the galaxy group members. We compare the SFRs depending on galaxy proximity to the nebulae and QSO. 

\section{Methods} 




The PKS0405$-$123 field was observed with the Multi Unit Spectroscopic Explorer (MUSE, \citealt{2010SPIE.7735E..08B}), an integral field spectrograph (IFS) on the Very Large Telescope (VLT; PI: J. Schaye, PID: 094.A-0131) with coverage from 4650-9300\AA\ at $R$=2000$-$4000. The data were reduced using the standard MUSE reduction pipeline \citep{weilbacher2020} and with tools from the {\tt CubExtractor} package \citep{cantalupo2019}. Galaxies in the field were identified using {\tt Source Extractor} \citep{Bertin_1996}, and spectral extractions were performed with {\tt PyMUSE} \citep{pessa2020}. Redshifts for the galaxies were determined with the spectral template fitting code {\tt REDROCK} \citep{ross2020}. Two galaxy spectra (G1 and G2) were corrected by subtracting off extraneous light from the QSO; we extracted a 2\farcs5 circular region at the center of the QSO and normalized the spectra to the continuum around the [OII]$\lambda$$\lambda$3727,3729 doublet. Through visual inspection of the strong optical lines, we categorized each galaxy as either emission-line or absorption-line dominated.

We used the [OII]$\lambda$$\lambda$3727,3729 emission line to characterize the galaxy SFRs. We fit double Gaussian curves in order to get averaged [OII] flux values for each galaxy. Redshifts from {\tt REDROCK} were used in the calculations to determine the location of the mean of each [OII] doublet. We used equations from \citet{Moustakas_2006} to calculate the [OII] SFRs, which depend on luminosity and Vega \textit{B} magnitudes. Using {\tt Source Extractor}, we determined the HST/ACS F814W galaxy magnitudes and converted them to Vega \textit{B} magnitudes. In order to take cosmological effects due to redshift into account in the SFR calculations, k-corrections were made based off of methods shown in \citet{k-correct}. We produced SFRs for each of the 30 galaxies. Four galaxies in the sample had negligible [OII] emission lines. We report their SFRs as 2$\sigma$ upper limits.

\begin{figure}
\plotone{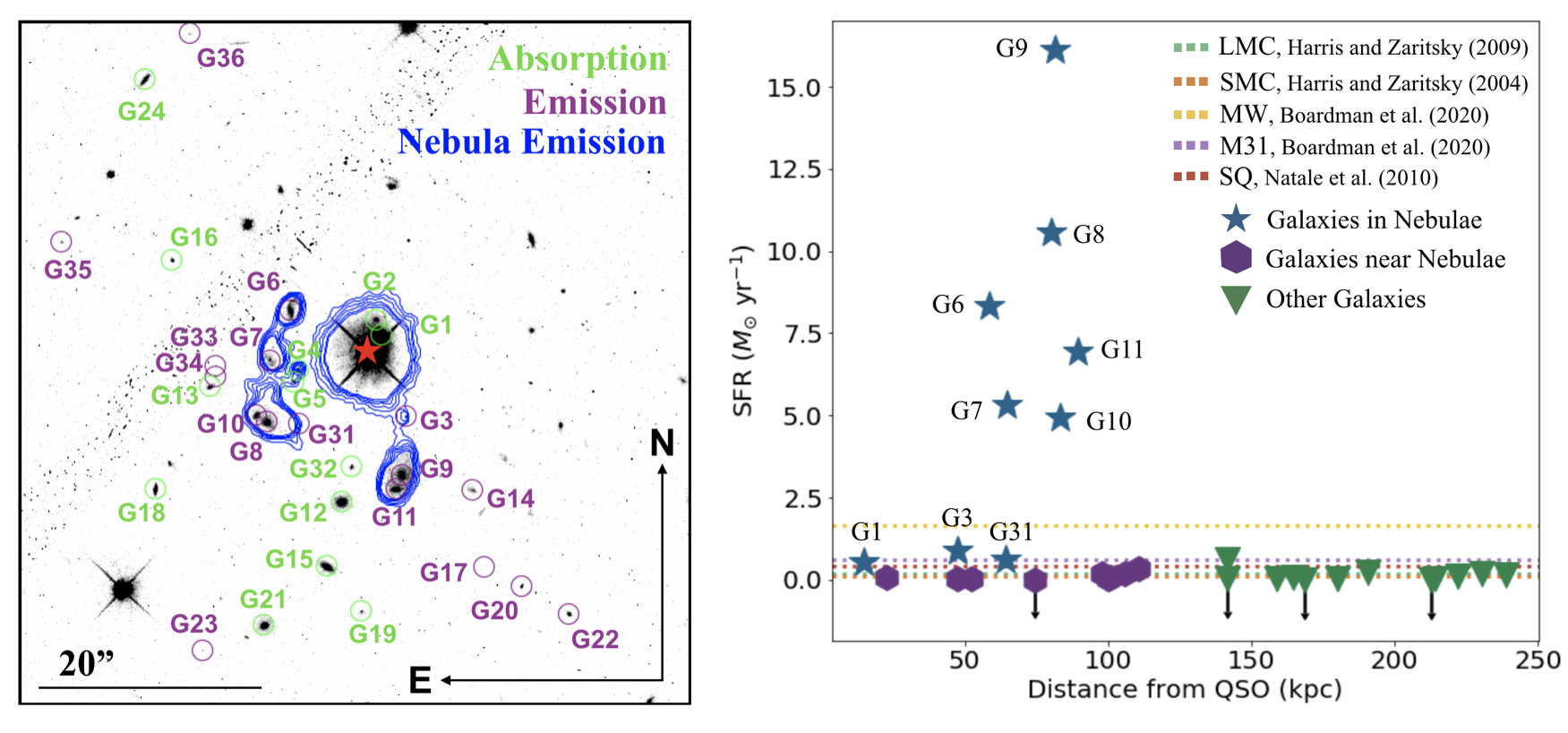}
\caption{Left: HST/ACS F814W filter image of the PKS0405$-$123 field. The blue contours signify [OII]$\lambda$$\lambda$3727,3729 nebulae emission, and the red star signifies the QSO. Right: [OII] SFR versus distance from QSO for the galaxy sample compared to other well known galaxies. The six galaxies with significantly higher SFRs, each within nebulae, are made up of three pairs: G9 and G11, G8 and G10, G6 and G7. 
\label{fig:general}}
\end{figure}

\section{Results} 

Thirty galaxies are found within $|\Delta v| <$ 2,000 \kms\ of the QSO redshift ($z_{\rm QSO} = 0.5731 \pm 0.0003$; \citealt{johnson2018}), which we refer to as members of the galaxy group that houses QSO PKS0405$-$123. The left plot in Figure~\ref{fig:general} presents an HST/ACS F814W image of the field with the host QSO, our sample of galaxies, and the [OII] contours of the nebulae from the MUSE observations overlaid. 
Emission-line dominated galaxies make up 57$\%$ of our sample, and the remaining 43$\%$ of the sample is comprised of absorption-line dominated galaxies. Nine of the galaxies are within nebulae, nine are close to nebulae, and the other twelve are not in or near nebulae. The galaxy in our sample with the highest SFR is G9 with an SFR = $16.11\pm 0.06$ \msunyr. 

We consider several different average SFR groupings based on galaxy location and activity. We refer to galaxies with SFRs above 1 \msunyr\ as galaxies with active star formation. The average SFR of active star-forming galaxies is SFR = $8.70\pm 0.02$ \msunyr. The average SFR of inactive galaxies is SFR = $0.19\pm 0.07$ \msunyr. When we characterize by proximity to the nebulae, we find the average SFR of galaxies located within the nebulae is SFR = $6.03\pm 0.01$ \msunyr. All of the galaxies located within nebulae have active star formation except for G1, the galaxy closest to the QSO. The average SFR of galaxies located close to but not within nebulae is SFR = $0.11\pm 0.11$ \msunyr, defined as a galaxy within 5$\arcsec$ of any part of a nebula. The average SFR of all galaxies not in or close to nebulae is SFR = $0.12\pm 0.12$ \msunyr. On average, galaxies located inside nebulae have significantly higher SFRs than those outside of nebulae. 

The right plot in Figure~\ref{fig:general} presents the SFR for each galaxy versus relative distances to the QSO. 
The plot shows no evidence of possible correlation between distance from the QSO and SFR. This suggests that the QSO is not driving the higher SFRs in the sample. The plot additionally compares SFRs in our group to those of other galaxies and galaxy groups: the SFRs of the Milky Way and M31, 
 the average SFR of four galaxies in Stephan's Quintet, 
and the SFRs of the Small and Large Magellanic Clouds. 
The majority of the SFRs calculated for our sample are consistent with the SFRs of the comparison galaxies and galaxy groups. 

There are three close pairs of galaxies (G9 and G11, G8 and G10, G6 and G7) with significantly enhanced SFRs, all of which are within nebulae.
Figure 2 of \citet{johnson2018} presents a velocity space plot, which shows that the galaxies within the nebulae tend to have the same velocities as the nebulae. Given this information, we conclude that the three large gaseous nebulae were created by merger interactions that are enhancing the SFRs of the involved galaxies. We do not observe any statistically significant influence of the QSO on the SFRs. This is in agreement with the findings of \citet{johnson2018}.

\bibliography{sample631}{}

\end{document}